\documentclass[aps,prc,preprint,showpacs]{revtex4-1}  
\usepackage{graphicx}
\usepackage{amsmath}

\newcommand {\nc} {\newcommand}
\nc {\beq} {\begin{eqnarray}} \nc {\eol} {\nonumber \\} \nc {\eeq}
{\end{eqnarray}} \nc {\eeqn} [1] {\label{#1} \end{eqnarray}} \nc
{\eoln} [1] {\label{#1} \\} \nc {\ve} [1] {\mbox{\boldmath $#1$}}
\nc {\rref} [1] {(\ref{#1})} \nc {\Eq} [1] {Eq.~(\ref{#1})} \nc
{\re} [1] {Ref.~\cite{#1}} \nc {\dem} {\mbox{$\frac{1}{2}$}} \nc
{\arrow} [2] {\mbox{$\mathop{\rightarrow}\limits_{#1 \rightarrow
#2}$}}
\begin{document}
\title{Astrophysical  $^{3}{\rm He}(\alpha, \gamma)^{7}{\rm Be}$
and $^{3}{\rm H}(\alpha,\gamma)^{7}{\rm Li}$ direct capture
reactions in a potential model approach}

\author {E. M. Tursunov}
\email{tursune@inp.uz} \affiliation {Institute of Nuclear Physics,
Academy of Sciences, 100214, Ulugbek, Tashkent, Uzbekistan}
\author {S. A. Turakulov}
\email{turakulov@inp.uz} \affiliation {Institute of Nuclear Physics,
Academy of Sciences, 100214, Ulugbek, Tashkent, Uzbekistan}
\author {A. S. Kadyrov}
\email{a.kadyrov@curtin.edu.au} \affiliation{Curtin Institute for Computation and Department of Physics
and Astronomy, Curtin University, GPO Box U1987, Perth, WA 6845,
Australia}

\begin{abstract}
The astrophysical $^{3}{\rm He}(\alpha, \gamma)^{7}{\rm Be}$ and
$^{3}{\rm H}(\alpha, \gamma)^{7}{\rm Li}$ direct capture processes
are studied in the framework of the two-body model with the
potentials of a simple Gaussian form, which describe correctly the
phase-shifts in the s-, p-, d-, and f-waves, as well as the binding
energy and the asymptotic normalization constant of the ground
$p_{3/2}$ and the first excited $p_{1/2}$ bound states. It is shown
that the E1-transition from the initial s-wave to the final p-waves
is strongly dominant in both capture reactions. On this basis the
s-wave potential parameters are adjusted to reproduce the new data
of the LUNA collaboration around 100 keV and the newest data at the
Gamov peak estimated with the help of the observed neutrino fluxes
from the Sun, $S_{34}$(23$^{+6}_{-5}$ keV)=0.548$\pm$0.054 keV b for
the astrophysical S-factor of the capture process $^{3}{\rm
He}(\alpha, \gamma)^{7}{\rm Be}$. The resulting model describes well
the astrophysical S-factor in low-energy Big Bang nucleosynthesis
region of 180-400 keV, however has a tendency to underestimate the
data above 0.5 MeV. The energy dependence of the S-factor is
mostly consistent with the data and the results of the no-core shell
model with continuum, but substantially different from the fermionic
molecular dynamics model predictions. Two-body potentials, adjusted
on the properties of the $^7$Be nucleus, $^3{\rm He}+\alpha$ elastic
scattering data and the astrophysical S-factor of the $^{3}{\rm
He}(\alpha, \gamma)^{7}{\rm Be}$ direct capture reaction, are able
to reproduce the properties of the $^7$Li nucleus, the binding
energies of the ground 3/2$^-$ and first excited 1/2$^-$ states, and
phase shifts of the $^3 {\rm H}+\alpha$ elastic scattering in
partial waves. Most importantly, these potential models can
successfully describe both absolute value and energy dependence of
the existing experimental data for the mirror astrophysical
$^{3}{\rm H}(\alpha, \gamma)^{7}{\rm Li}$ capture reaction without
any additional adjustment of the parameters.
\end{abstract}


 \maketitle

\section{Introduction}
The radiative capture   $^{3}{\rm He}(\alpha, \gamma)^{7}{\rm Be}$
and $^{3}{\rm H}(\alpha, \gamma)^{7}{\rm Li}$ processes are the key
nuclear reactions in stellar nucleosynthesis
\cite{adelber11,fields11}. Both of these reactions are important for
studies of the primordial nucleosynthesis, in particular, for the
solution of the so-called $^{7}{\rm Li}$ abundance problem
\cite{asp06}. In addition, the $^{3}{\rm He}(\alpha, \gamma)^{7}{\rm
Be}$ reaction is very useful for the study of the kinetics of
processes taking place in the Sun since it is a starting point for
the second and third chains in the $pp$-cycle of hydrogen burning.
On the other hand the $^{7}{\rm Be}$ nucleus plays a dominant role
in the neutrino production processes in both solar and Big Bang
nucleosynthesis (BBN) models.

Experimental studies of the $^{3}{\rm He}(\alpha, \gamma)^{7}{\rm
Be}$ and $^{3}{\rm H}(\alpha, \gamma)^{7}{\rm Li}$ radiative capture
processes  started in 1960s \cite{grif61,parker63}. Since then these
reactions have consistently attracted interest of experimentalists
\cite{kraw82,osborn82,hilg88,schr87,burz87,brune94}.  Recent
measurements were reported in
\cite{Nara04,bem06,confort07,brown07,leva09,carmona12,bordeanu13,carmona14}.
The main difficulty in laboratory studies of these processes at low
energies of astrophysical relevance (roughly from 20 to 500 keV) is
related with the presence of strong Coulomb repulsive forces,
especially for the production of the $^7{\rm Be}$ nucleus. Due to
this difficulty the measured values of the astrophysical S-factor
contain large uncertainties. The most accurate experimental results
for the astrophysical S-factor were obtained by the LUNA
collaboration \cite{bem06,confort07} in a low-energy region around
$E_{\rm cm}=$100 keV, where $E_{\rm cm}$ is the collision energy in
the center of mass (cm) frame. The experimental uncertainties in the
measured values of the astrophysical S-factor around 70 keV b are
much smaller than those in the old data. However, even these smaller
error bars can have a strong influence on estimations of the
astrophysical reaction rates in the BBN and solar models
\cite{adelber11}. Therefore, there is still a need for more accurate
experimental studies in the low energy region.

Very recently observed neutrino fluxes from the Sun were used to
estimate the $^{3}{\rm He}(\alpha, \gamma)^{7}{\rm Be}$ astrophysical S-factor
within the standard solar model at the Gamow peak to be
$S_{34}$(23$^{+6}_{-5}$ keV)=0.548$\pm$0.054 keV b \cite{takacs15}.
This new data point was then used for evaluation of the
astrophysical S-factor at Big Bang energies and the corresponding
thermonuclear reaction rates. However, an estimate of the
primordial lithium abundance, $^7$Li/H=5$\times$ 10$^{-10}$, obtained
in the model is much larger than the observed Spite plateau
\cite{fields11}.

From the theoretical side, potential models
\cite{mohr09,mason09,dub10}, microscopic R-matrix approach
\cite{desc10}, microscopic cluster models \cite{kajino87,vasil12},
microscopic approach based on an algebraic version of the resonating
group method \cite{sol14}, fermionic molecular dynamics (FMD) method
\cite{neff11}, no-core shell model with continuum (NCSMC)
\cite{doh16} and the semimicroscopic phenomenological approach
\cite{noll01} have been developed to study  the astrophysical
$^{3}{\rm He}(\alpha, \gamma)^{7}{\rm Be}$ and $^{3}{\rm H}(\alpha,
\gamma)^{7}{\rm Li}$ reactions. The most elaborate microscopic
approaches based on the NCSMC and FMD yield an overall good
description of the experimental data except the old data from Ref.
\cite{parker63} which are now believed to be less accurate. However
the astrophysical S-factor obtained within these two methods show
different energy dependence for both capture processes. At the same
time they describe well the data of the LUNA collaboration
\cite{bem06,confort07} and the newest data coming from the observed
neutrino flux at the Gamov peak \cite{takacs15}. One should note
that a fully ab-initio calculation of these radiative capture
reactions, including three-body nuclear forces is not yet available
and is still a big challenge. On the other hand, the question
whether or not a simple potential model is able to reproduce the
available data for the capture reactions at least in the BBN energy
region of $E_{\rm cm}$=180-400 keV remains to be answered. How does
the description of the data for the astrophysical S-factor compare
with the corresponding ab-initio results? To our best knowledge,
these questions are still open. The most realistic potential model
\cite{mohr09} based on folding potentials agrees well with the old
data \cite{parker63}, which is much lower than the new data from
Refs.\cite{bem06,confort07} at the astrophysical, low energy region.

Potential cluster models are able to reproduce both the bound state
properties and the scattering data \cite{dub10,fraser}. An important
feature of the potential models is that the two-body potentials have
to be adjusted to reproduce not only the phase shifts in all partial
waves and the binding energies of the bound states, but also the
asymptotic properties of the bound state wave functions, like the
asymptotic normalization coefficient (ANC). The importance of
asymptotic properties of the two-body potentials have been
demonstrated for the astrophysical  $\alpha(d,\gamma)^6{\rm Li}$
capture process at low energies \cite{tur15,tur16}.
Required empirical values of the asymptotic normalization
coefficient can be extracted from the scattering data within
different approaches, e.g. analytic continuation to the S-matrix
pole \cite{blokh93}, the effective range method
\cite{spar10,blokh17,blokh18} and distorted-wave Born approximation (DWBA)
\cite{olim16}.

The potential models can also be used to improve the
accuracy of the direct experiments on astrophysical capture
reactions. Recently, a photon angular distribution calculated in the
potential model has been used \cite{mukh16} to find the best
kinematic conditions for the measurement of the
$^2$H$(\alpha,\gamma)^6$Li reaction.

The aim of present paper is to study in detail the astrophysical
 $^{3}{\rm He}(\alpha, \gamma)^{7}{\rm Be}$ and $^{3}{\rm H}(\alpha,
\gamma)^{7}{\rm Li}$ capture reactions in a potential model. As it
is known from the literature, and as will be seen below, the most
important contribution to above processes at low astrophysical
energies comes from the dipole E1-transition operator, while the
E2-transition only gives a small contribution in the resonance
energy region. The M1-transition is also strongly suppressed. The
two-body Gaussian potentials \cite{dub10} which reproduce the bound
states energies and the phase shifts in each partial wave will be
examined. The potential parameters will be adjusted to reproduce the
empirical values of the asymptotic normalization coefficient in the
$p_{3/2}$- and $p_{1/2}$-bound states of the $^7$Be nucleus,
recently extracted from the phase-shift analysis within the DWBA
method \cite{olim16} and from the analysis of the experimental
S-factor \cite{qah12}. The d- and f-wave potentials from
Ref.~\cite{dub10}, which describe  the corresponding phase shifts
well, will be applied.

As the E1-transition occurs from the initial s-wave scattering state
to the final p-wave bound states, the choice of the s-wave
potentials is the next most important point of the potential model.
The existence of infinite number of phase-equivalent potentials
opens an unique possibility to adjust the S-wave potential
parameters to the experimental astrophysical S-factor. The nodal
positions of the s-wave scattering wave function, as well as the
p-wave bound state wave functions at short distances due to their
orthogonality to the Pauli forbidden states (two in s-wave and one
in each of the partial $p_{1/2}$ and $p_{3/2}$-waves) play a crucial
role in decreasing effective overlap integrals, involving these two
wave functions, thus resulting in the low values of the
astrophysical S-factor, consistent with the experimental results. In
this sense a role of the Pauli forbidden states in the capture
process is similar to that in the beta-decay of the $^6${\rm He}
halo nucleus into the $\alpha-d$ continuum \cite{tur06,tur06a}.

At the first step  the initial potential from
Ref. \cite{dub10} will be examined in the s-wave. After that we will show that it
is possible to find the most suitable model among the
phase-equivalent potentials, fitting the s-wave potential parameters
to the astrophysical S-factor of the LUNA collaboration
\cite{bem06,confort07} and the newest data at the Gamov peak
\cite{takacs15} in low energy region.
The astrophysical S-factor of the mirror reaction $^{3}{\rm
H}(\alpha, \gamma)^{7}{\rm Li}$ will be estimated with the same
potentials, constructed from the study of the $^{3}{\rm He}(\alpha,
\gamma)^{7}{\rm Be}$ capture process by appropriate modification of
the Coulomb interaction potential due to the difference in the
charge values of the clusters $^3$He and $^3$H.

The theoretical model will be briefly described in Section II,
numerical results will be given in Section III, and conclusions will
be drawn in the last section.

\section{Theoretical model}
\subsection{Wave functions}

 In a single channel approximation, the initial and final state wave functions are defined as
\begin{eqnarray}
\Psi_{lS}^{J}=\frac{u_E^{(lSJ)}(r)}{r}\left\{Y_{l}(\hat{r})\otimes\chi_{S}(\xi)
\right\}_{J M}
\end{eqnarray}
and
\begin{eqnarray}
\Psi_{l_f S}^{J_f}
=\frac{u^{(l_fSJ_f)}(r)}{r}\left\{Y_{l_f}(\hat{r})\otimes\chi_{S}(\xi)
\right\}_{J_f M_f},
\end{eqnarray}
respectively. The radial wave functions of the initial $\alpha-^{3}$He and
$\alpha-^{3}$H scattering states
 in the  $s_{1/2}$, $p_{1/2}$, $p_{3/2}$, $d_{3/2}$, $d_{5/2}$, $f_{5/2}$, $f_{7/2} $  partial
waves are found as solutions of the two-body Schr\"{o}dinger
equation
\begin{eqnarray}
\left[-\frac{\hbar^2}{2\mu}\left(\frac{d^2}{dr^2}-\frac{l(l+1)}{r^2}\right)+V^{
lSJ}(r)\right] u_E^{(lSJ)}(r)=E u_E^{(lSJ)}(r),
\end{eqnarray}
where $\mu$ is the reduced mass of the clusters involved in the
capture process, ${1}/{\mu}={1}/{m_1}+{1}/{m_2}$ , and
$V^{lSJ}(r)$ is a two-body potential in the partial wave with the
orbital momentum $l$, spin $S$ and total momentum $J$. The wave
functions $u^{(l_fSJ_f)}(r)$ of the final $p_{3/2}$ ground and
$p_{1/2}$ excited bound states are found as solutions of the
bound-state Schr\"{o}dinger equation. For the solution of the
Schr\"{o}dinger equation the Numerov algorithm of a high accuracy of
order $O(h^{6})$ is applied. The calculated wave functions allow one
to estimate the characteristics of the astrophysical capture
reactions $^{3}{\rm He}(\alpha, \gamma)^{7}{\rm Be}$ and $^{3}{\rm
H}(\alpha, \gamma)^{7}{\rm Li}$, the cross section and the
astrophysical S-factor.

 The radial scattering wave function is normalized  with the help of the
 asymptotic relation
  \beq u_E^{(lSJ)}(r) \arrow{r}{\infty} \cos\delta_{lSJ}(E) F_l (\eta,kr)
+ \sin\delta_{lSJ}(E) G_l(\eta,kr), \label{eq220} \eeq where $k$ is
the wave number of the relative motion, $\eta$ is the
Zommerfeld parameter, $F_l$ and $G_l$ are regular and irregular
Coulomb functions, respectively, and $\delta_{lSJ}(E)$ is the phase
shift in the $(l,S,J)$th partial wave.

The $\alpha-^3$He and $\alpha-^3$H two-body potentials are taken in
a simple Gaussian form \cite{dub10}:
 \beq
 V^{lSJ}(r)=V_0 \exp(-\alpha r^2)+V_c(r),
\label{pot}
 \eeq
 where the Coulomb part is given as
\begin{eqnarray}
 V_c(r)=
\left\{
\begin{array}{lc}
 Z_1 Z_2 e^2/r &  {\rm if} \,\, r>R_c, \\
Z_1 Z_2 e^2 \left(3-{r^2}/{R_c^2}\right)/(2R_c) &  {\rm otherwise},
\end{array}
\right. \label{Coulomb}
\end{eqnarray}
with the Coulomb parameter $R_c$, and charge numbers $Z_1$, $Z_2$ of
the first and second clusters, respectively. The parameters
$\alpha$, $V_0$ and $R_c$ of the potential are specified for each
partial wave.

\subsection{Cross sections of the radiative capture process}
The cross sections of the radiative capture process  read
\cite{angulo, dub10}
\begin{eqnarray}
\sigma(E)=\sum_{J_f \lambda \Omega}\sigma_{J_f \lambda}(\Omega),
\end{eqnarray}
where $\Omega=$ E  or M (electric or magnetic transition), $\lambda$ is a multiplicity of the transition, $J_f$ is the
total angular momentum of the final state. For a particular final
state with total momentum $J_f$ and multiplicity $\lambda$ we have
\begin{align}
 \sigma_{J_f \lambda}(\Omega) =& \sum_{J}\frac{(2J_f+1)} {\left
[S_1 \right]\left[S_2\right]} \frac{32 \pi^2 (\lambda+1)}{\hbar
\lambda \left( \left[ \lambda
\right]!! \right)^2} k_{\gamma}^{2 \lambda+1} C^2(S) \nonumber \\
&\times \sum_{l S}
 \frac{1}{ k_i^2 v_i}\mid
 \langle \Psi_{l_f S}^{J_f}
\|M_\lambda^\Omega\| \Psi_{l S}^{J} \rangle \mid^2,
\end{align}
where $l,l_{f}$ are the orbital momenta of the initial and final
states, respectively, $k_i$ and $v_i$ are the wave number and
velocity of the $\alpha-^3$He (or $\alpha-^3$H) relative motion of
the entrance channel, respectively; $S_1$, $S_2$ are spins of the
clusters $\alpha$ and $^3${\rm He} (or $^3$H), $k_{\gamma}=E_\gamma
/ \hbar c$ is the wave number of the photon corresponding to energy
$E_\gamma=E_{\rm th}+E$, where $E_{\rm th}$ is the threshold energy.
Constant $C^2(S)$ is the spectroscopic factor \cite{angulo}. As it
was argued in Ref.\cite{mukh11}, within the potential approach its
value must be taken equal to 1 if the phase shifts in the partial
waves are correctly reproduced. We also use short-hand notations
$[S]=2S+1$ and $[\lambda]!!=(2\lambda+1)!!$.

The reduced matrix elements are evaluated between the initial and
final states represented by wave functions $\Psi_{l S}^{J}$ and
$\Psi_{l_f S}^{J_f}$, respectively.

The electric transition operator in the long-wavelength
approximation reads
\begin{eqnarray}
M_{\lambda m}^E=e \sum_{j=1}^{A} Z_j{r'_j}^{\lambda}Y_{\lambda
m}(\hat{r'}_j),
\end{eqnarray}
where $\vec {r'}_{j}= \vec{r}_j-\vec{R}_{cm}$ is the radius vector
of the $j$th particle in the center of mass system. Its reduced matrix
elements can be evaluated as follows:
\begin{eqnarray}
\langle \Psi_{l_f S}^{J_f}\|M_\lambda^E\| \Psi_{l S}^{J} \rangle
 &=& e\left[Z_1 \left( \frac{A_2}{A} \right)^{\lambda}+Z_2
\left(\frac{-A_1}{A} \right)^{\lambda} \right]      \\
\nonumber && \times (-1)^{J+l+S}\left(\frac{[\lambda][l][J]}{4
\pi}\right)^{1/2} C^{l_f 0}_{\lambda 0 l 0} \left\{
\begin{array}{ccc}
J & l & S \\
l_{f} & J_{f} & \lambda
\end{array} \right\} \int^{\infty}_{0} u_{E}^{(lSJ)}(r)r^{\lambda} u^{(l_fSJ_f)} (r) dr,
\end{eqnarray}
where $A_1$, $A_2$  are mass numbers of the clusters in the entrance
channel, $A=A_1+A_2$.

The magnetic transition operator reads
\begin{eqnarray}
M_{1 \mu}^M&=& \sqrt{\frac {3}{4 \pi}}\left[ \sum_{j=1}^{A}
 \mu_N \frac{Z_j}{A_j}\hat{l}_{j \mu} + 2 \mu_j \hat{S}_{j \mu} \right]
 \nonumber \\
 &=& \sqrt{\frac {3}{4 \pi}} \left[\mu_N \left( \frac{A_2 Z_1}{A A_1} + \frac{A_1
 Z_2}{A A_2} \right) \hat{l}_{r \mu} +2(\mu_1\hat{S}_{1\mu}+
 \mu_2\hat{S}_{2\mu})\right],
\end{eqnarray}
where $\mu_N$ is the nuclear magneton, $\mu_j$ is the magnetic
moment and $\hat{l}_{j \mu}$ is the orbital momentum of $j$th particle. The angular momentum of the relative
motion is denoted as $\hat{l}_{r \mu}$.
 The reduced matrix elements of the magnetic M1 transition
operator can be evaluated as
\begin{eqnarray}
\langle \Psi_{l_f S}^{J_f}\|M_1^M\| \Psi_{l S}^{J} \rangle & = &
\mu_N \left( \frac{A_2 Z_1}{A A_1} + \frac{A_1
 Z_2}{A A_2} \right) \sqrt{l_f(l_f+1)[J_f][l_f]}(-1)^{S+1+J_f+l_f}
\left\{
\begin{array}{ccc}
l_f & S & J_f \\
J & 1 & l_f
\end{array}
\right\} I_{if}  \nonumber   \\
&&+  2 \mu(^3{\rm He})(-1)^{1+l_f+3S-J} \sqrt{S(S+1)[S][J_f]}
\left\{
\begin{array}{ccc}
S & l_f & J_f \\
J & 1 & S
\end{array}
\right\}  I_{if},
\end{eqnarray}
where the overlap integral is given as
\begin{eqnarray}
I_{if}= \delta_{l l_f} \sqrt{\frac {3}{4 \pi}} \int^{\infty}_{0}
u_{E}^{(lSJ)}(r)u^{(l_fSJ_f)} (r) dr.
\end{eqnarray}
In addition, $\mu(^3{\rm He})=$-2.1275 $\mu_N$ is the magnetic
momentum of the $^3$He nucleus, which must be replaced by the
magnetic momentum $\mu(^3{\rm H})=$ 2.979 $\mu_N$ of the $^3$H
nucleus for the mirror reaction.

Finally, the astrophysical $S$-factor of the process is expressed in terms of the
cross section as \cite{Fowler}
\begin{eqnarray}
S(E)=E \, \, \sigma(E) \exp(2 \pi \eta).
\end{eqnarray}

\section{Numerical results}
\subsection{Details of the calculations and phase-shift descriptions}

\par For the solution of the Schr\"{o}dinger equation in the entrance
and exit channels we use the two-body $\alpha-^{3}${\rm He} and
$\alpha-^{3}$H central potentials of the Gaussian form from Ref.
\cite{dub10} as defined in Eq.(\ref{pot}) with the corresponding
Coulomb part, see Eq.(\ref{Coulomb}). For consistency we use
the same parameters as in the aforementioned paper. Namely, we use $\hbar^2/2$[a.m.u]=20.7343 MeV fm$^2$ and the Coulomb parameter $R_c$=3.095 fm.
The experimental mass values are also taken from Ref. \cite{dub10}:
$m_{^4{\rm He}}=A_1$ a.m.u. $=$4.001506179127 a.m.u., $m_{^3{\rm
He}}=A_2$ a.m.u. $=$ 3.0149322473 a.m.u. or $m_{^3{\rm H}}=A_2$
a.m.u. $=$ 3.0155007134 a.m.u.

The scattering wave function $u_{E}(r)$ of the relative motion is
calculated as a solution of the Schr{\"o}dinger equation using the
Numerov method with an appropriate potential subject to the boundary
condition specified in Eq.(\ref{eq220}).

The depth parameters $V_0$ of the $\alpha-^{3}$He and $\alpha-^{3}$H
potentials are given in Table \ref{tab03}. All the presented potentials,
including the initial deep potential $V_D$ of Dubovichenko
\cite{dub10} reproduce the experimental phase shifts of
$\alpha-^{3}${\rm He} scattering in all the partial waves and the binding
energies $E_b(3/2^-)$=1.5866 MeV and $E_b(1/2^-)$=1.16082 MeV of the
$^7$Be nucleus bound states.

The width parameter of the initial potential $V_D$ \cite{dub10} was
chosen as $\alpha=$0.15747 fm$^{-2}$ for all the partial waves. It
yields the ANC values $C(3/2^-)$=4.34 fm$^{-1/2}$ and
$C(1/2^-)$=3.71 fm$^{-1/2}$  for the bound states.

The parameters of the modified potential $V_{M1}$ in the p-waves are
fitted to reproduce the empirical values of ANC, extracted from the
experimental $\alpha-^3${\rm He} scattering data given in
Ref.~\cite{olim16} within the DWBA method, $C(3/2^-)$=4.785
fm$^{-1/2}$ and $C(1/2^-)$= 4.243 fm$^{-1/2}$, while keeping the
experimental phase-shift description and the binding energies. The
values of the depth parameter and the width parameter $\alpha$
for the $p_{3/2}$ and $p_{1/2}$ bound states are given in Tables \ref{tab03} and \ref{tab04}, respectively. The
parameters in other waves are identical to that of $V_D$.
\begin{figure}[htb]
\includegraphics[width=178mm]{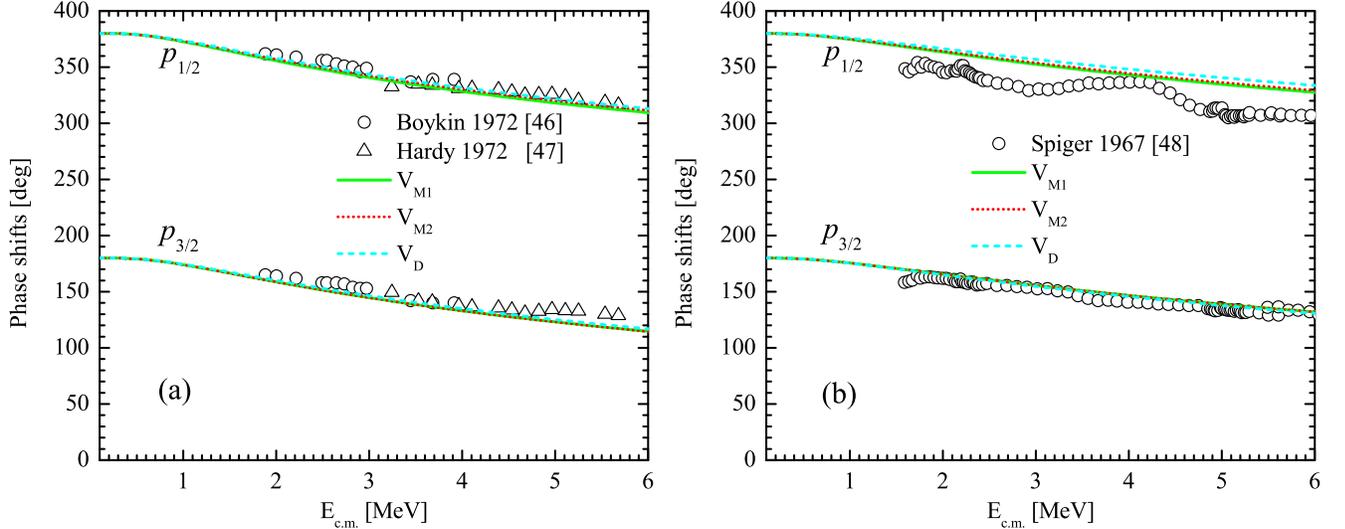}
\caption{p-wave phase-shift description of the $^{3}{\rm He}+\alpha$
and $^{3}{\rm H}+\alpha$ scattering within different potential
models in comparison with available data.} \label{FIG1}
\end{figure}

 In addition to the phase shifts and the binding energies, the
modified potential $V_{M2}$ is adjusted to reproduce  the empirical
values of ANC, extracted from the analysis of the experimental
astrophysical S-factor of the $\alpha(^3${\rm He},$\gamma$)$^7$Be
capture reaction presented in Ref.\cite{qah12}, $C(3/2^-)$=4.80
fm$^{-1/2}$ and $C(1/2^-)$= 3.94 fm$^{-1/2}$. The values of $\alpha$
for the $p_{3/2}$ and $p_{1/2}$ bound states are given in Table
\ref{tab04}. Again the potential parameters in all the partial
waves, except p-waves, are the same as in $V_D$ and $V_{M1}$.

The calculated p-wave phase shifts for the $^3$He$+\alpha$ and
$^3$H$+\alpha$ scattering are shown on Fig. \ref{FIG1} in comparison
with experimental data from Refs.\cite{boykin,hardy,spiger}. As can
be seen from the figure, the modified potentials $V_{M1}$ and
$V_{M2}$ yield equally good phase-shift description as the initial
potential $V_D.$

As we see below, the potentials $V_D$, $V_{M1}$ and $V_{M2}$ do not
reproduce the new data of the LUNA collaboration
\cite{bem06,confort07} at energies around 100 keV and the newest
data \cite{takacs15} at the Gamov peak  $S_{34}$(23$^{+6}_{-5}$
keV)=0.548$\pm$0.054 keV b for the astrophysical S-factor of the
$^{3}{\rm He}(\alpha, \gamma)^{7}{\rm Be}$ capture reaction. The
unique property of the potential model is that there is a
possibility to adjust the potential parameters in the s-wave in
order to reproduce the new data for the astrophysical S-factor,
while keeping the experimental phase shifts unchanged. This is
possible because of the dominance of the E1-transition $s_{1/2} \to
p_{3/2}$ and $s_{1/2} \to p_{1/2}$ for the capture process. The
potential $V_{M1}^a$ is obtained as a modification of the $V_{M1}$
potential in the s-wave. Its depth and width parameter values are
given in Tables \ref{tab03} and \ref{tab04}, respectively. The
potential $V_{M2}^a$ was obtained from $V_{M2}$ in the same way. The
modified potentials $V_D^a$ and $V_D^b$ are built from the original
$V_D$ potential by the modification of the s-wave parameters. The
potentials $V_{M1}^a$, $V_{M2}^a$, and $V_D^a$ are adjusted to the
central value of the newest data at the Gamov peak, while the
$V_D^b$ is adjusted to the upper limit of the error bar of the
latter.
\begin{figure}[htb]
\includegraphics[width=178mm]{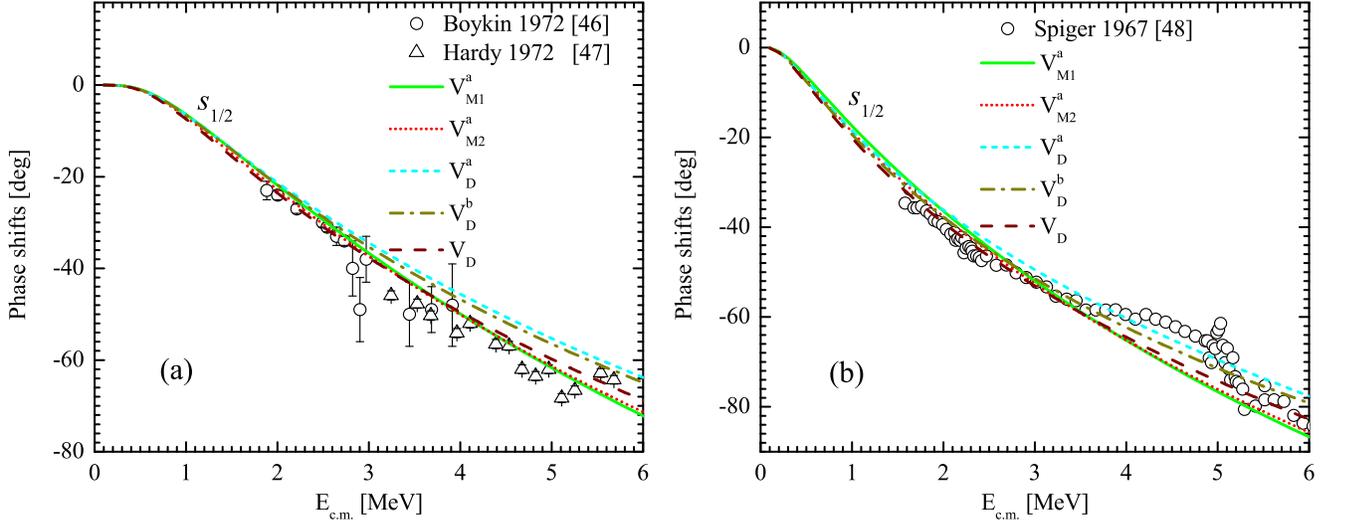}
\caption{s-wave phase-shift description of the $^{3}{\rm
He}+\alpha$ and $^{3}{\rm H}+\alpha$ scattering within different
potential models in comparison with available data.} \label{FIG2}
\end{figure}

The corresponding s-wave phase shift descriptions are shown in
Fig. \ref{FIG2} for the $^3$He$+\alpha$ and $^3$H$+\alpha$
scattering. In the energy range up to 3 MeV, the modified potentials
$V_{M1}^a$, $V_{M2}^a$, $V_D^a$ and $V_D^b$ describe the
experimental data at the level of the original potential $V_D$.
\begin{table}[htb]
\caption {Values of the depth parameter $V_0$, see  Eq. (\ref{pot}), of the
$\alpha - ^3${\rm He} ($^3$H) potential for different partial waves in
MeV.} \label{tab03}
{\small
\begin{tabular}{|c|c|c|c|c|c|c|c|c|c|}
\cline{2-3} \cline{4-5} \cline{6-7} \hline $L_J$& $V_D$ & $V_D^a$&
$V_D^b$ & $V_{M1}$ & $V_{M1}^a$ & $V_{M2}$ & $V_{M2}^a$ \\
\hline
  $s_{1/2}$ & $-67.5$ & $-77.0$ & $-130.0$ & -67.5 & $-50.0$ & -67.5 & -54.0   \\
   \hline \multicolumn{1}{|c|}{$p_{1/2}$} & \multicolumn{3}{|c|}{$-81.815179$ }&
   \multicolumn{2}{|c|}{$-70.912$} & \multicolumn{2}{|c|}{$-76.680$}\\
  \hline \multicolumn{1}{|c|}{$p_{3/2}$} &
  \multicolumn{3}{|c|}{$-83.589554$ }&\multicolumn{2}{|c|}{$-75.766$} &\multicolumn{2}{|c|}{$-75.486$}\\
  \hline \multicolumn{1}{|c|}{$d_{3/2}$} & \multicolumn{7}{|c|}{$-66.0$ } \\
  \hline \multicolumn{1}{|c|}{$d_{5/2}$} & \multicolumn{7}{|c|}{$-69.0$} \\
  \hline \multicolumn{1}{|c|}{$f_{5/2}$} & \multicolumn{7}{|c|}{$-75.9$} \\
  \hline \multicolumn{1}{|c|}{$f_{7/2}$} & \multicolumn{7}{|c|}{$-84.8$} \\
\hline
\end{tabular}
}
\end{table}
\begin{table}[htb]
\caption {Values of the width parameter $\alpha$, see  Eq.
(\ref{pot}), of the $\alpha - ^3${\rm He} ($^3$H) potential for
different partial waves in fm$^{-2}$.} \label{tab04} {\small
\begin{tabular}{|c|c|c|c|c|c|c|c|c|c|}
\cline{2-3} \cline{4-5} \cline{6-7} \hline $L_J$& $V_D$ & $V_D^a$&
$V_D^b$ & $V_{M1}$ & $V_{M1}^a$ & $V_{M2}$ & $V_{M2}^a$ \\
\hline
  $s_{1/2}$ & $0.15747$ & $0.180$ & $0.365$ & 0.15747& $0.109$ &0.15747 & 0.120  \\
   \hline \multicolumn{1}{|c|}{$p_{1/2}$} & \multicolumn{3}{|c|}{$0.15747$ }&
   \multicolumn{2}{|c|}{$0.1338$} & \multicolumn{2}{|c|}{$0.1463$}\\
  \hline \multicolumn{1}{|c|}{$p_{3/2}$} &
  \multicolumn{3}{|c|}{$0.15747$ }&\multicolumn{2}{|c|}{$0.1405$} &\multicolumn{2}{|c|}{$0.1399$}\\
  \hline \multicolumn{1}{|c|}{$d_{3/2}$} & \multicolumn{7}{|c|}{$0.15747$ } \\
  \hline \multicolumn{1}{|c|}{$d_{5/2}$} & \multicolumn{7}{|c|}{$0.15747$} \\
  \hline \multicolumn{1}{|c|}{$f_{5/2}$} & \multicolumn{7}{|c|}{$0.15747$} \\
  \hline \multicolumn{1}{|c|}{$f_{7/2}$} & \multicolumn{7}{|c|}{$0.15747$} \\
\hline
\end{tabular}
}
\end{table}

At first step a study of the astrophysical $^{3}{\rm He}(\alpha,
\gamma)^{7}{\rm Be}$ capture process will be performed within the
aforementioned  potential models. At the next step, these potentials
will be examined in the $^{3}$H$(\alpha, \gamma)^{7}$Li reaction
studies with the only modification of the Coulomb potential due to
different charge values of the $^3$He and $^3$H nuclei. The nuclear
part of the potentials will be kept unchanged on the basis of the
charge-independence property of nuclear forces.

\subsection{Estimation of the astrophysical S-factor for the $^{3}{\rm He}(\alpha,
\gamma)^{7}{\rm Be}$ capture process}

For the study of the $^{3}{\rm He}(\alpha, \gamma)^{7}{\rm Be}$
direct radiative capture process we first use the potentials $V_D$,
$V_{M1}$ and $V_{M2}$. As noted above, these potentials differ from
each other due to the parameters used in the $p_{1/2}$ and $p_{3/2}$
partial waves and yield different values for ANC. As mentioned above,
the $V_{M1}$ and $V_{M2}$ potentials were adjusted to the empirical
ANC values from Refs.~\cite{olim16} and \cite{qah12}, respectively.

\begin{figure}[htb]
\includegraphics[height=160mm]{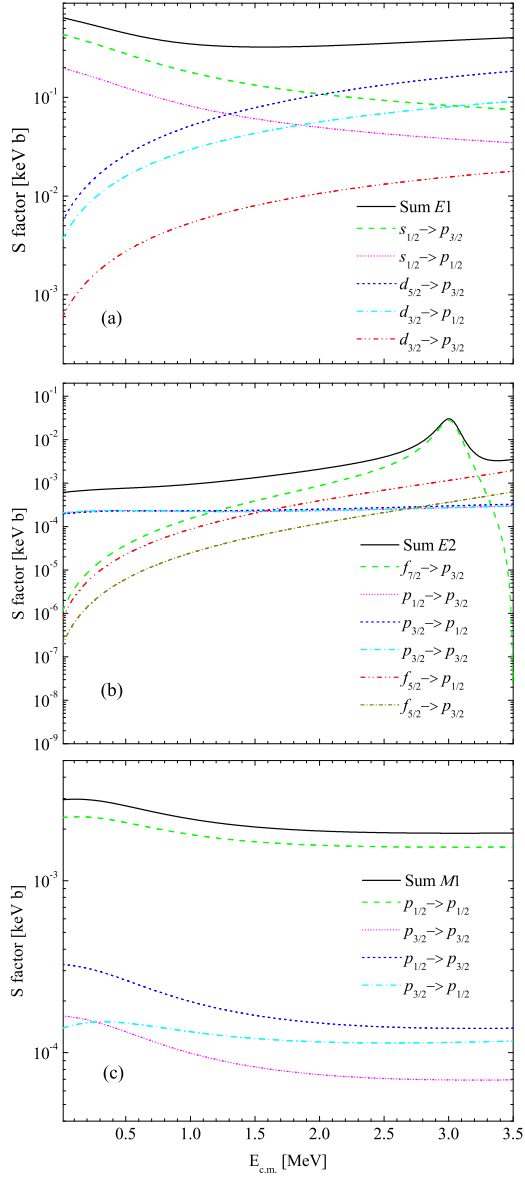}
\caption{Contributions of the partial $E$1-, $E$2- and
$M$1-components to the astrophysical $S$-factor for the $^{3}{\rm
He}(\alpha, \gamma)^{7}{\rm Be}$ capture process resulting from
calculations with the $V_{M1}$ potential. } \label{FIG3}
\end{figure}

\begin{figure}[htb]
\includegraphics[width=86mm]{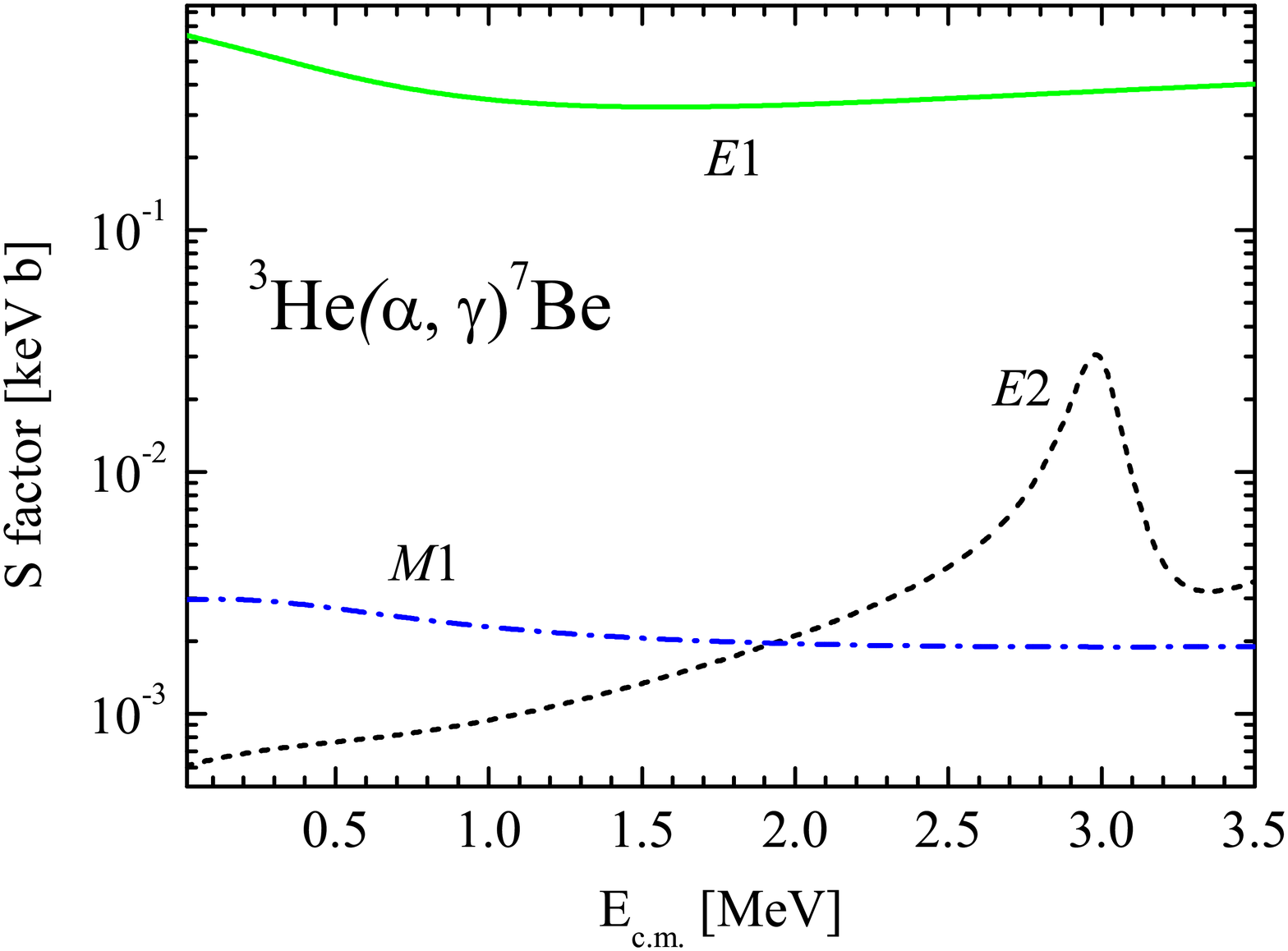}
\caption{Contributions of the total $E$1-, $E$2- and
$M$1-transitions to the astrophysical S-factor for the $^{3}{\rm
He}(\alpha, \gamma)^{7}{\rm Be}$ reaction estimated using the
$V_{M1}$ potential.} \label{FIG4}
\end{figure}

Contributions of the partial $E$1-transition components for the
$^{3}{\rm He}(\alpha, \gamma)^{7}{\rm Be}$ direct radiative capture
process are given in the left panel of Fig. \ref{FIG3} for the
$V_{M1}$ potential. As can be seen from the figure, the dominant
contribution in the astrophysical low energy region comes from the
E1-transition $s_{1/2} \to p_{3/2}$. The dominance is most
prevailing at energies close to zero. At energies above 2 MeV the
E1-transition from the $d_{5/2}$ to the $p_{3/2}$ partial wave
provides the largest contribution.
Contributions of the $E$2-components to the astrophysical $S$-factor
within the same $V_{M1}$ potential are shown in the middle panel of
Fig. \ref{FIG3}. The dominant contributions in low-energy region
correspond to the transitions between the p-waves. A resonance behavior
of the astrophysical S-factor at energies around 3 MeV is well
reproduced in the $f_{7/2} \to p_{3/2}$ transition.
Contributions of the partial $M$1-components to the astrophysical
$S$-factor for the $^{3}{\rm He}(\alpha, \gamma)^{7}{\rm Be}$ direct
capture process with the same $V_{M1}$ potential are displayed in
the third panel of Fig. \ref{FIG3}. Here the dominant contribution is
the M1-transition from the $p_{1/2}$ partial wave to the same one.

In order to compare the relative contributions from the electric
E1-, E2- and magnetic M1-transitions, in Fig. \ref{FIG4} we show the
summary of the results for the astrophysical S-factor of the
$^{3}{\rm He}(\alpha, \gamma)^{7}{\rm Be}$ capture reaction
calculated using the potential model $V_{M1}$. As can be seen from
the figure, the dominance of the E1-transition is maximal at the
zero energy where the contribution from the electric E1-transition
is larger than the sum of those from the E2- and M1-transitions by
more than two orders of magnitude.

\begin{figure}[htb]
\includegraphics[width=86mm]{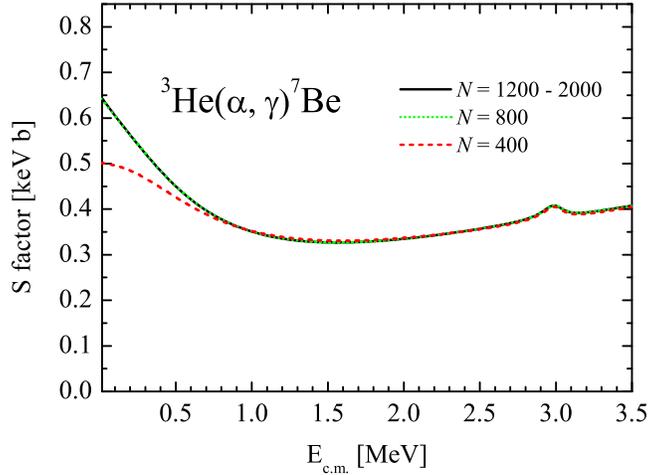}
\caption{Convergence of the astrophysical S-factor for the $^{3}{\rm
He}(\alpha, \gamma)^{7}{\rm Be}$ reaction with respect to the number
of integration points with fixed value of h=0.05 fm estimated with
the $V_{M1}$ potential.} \label{FIG5}
\end{figure}

\begin{figure}[htb]
\includegraphics[width=178mm]{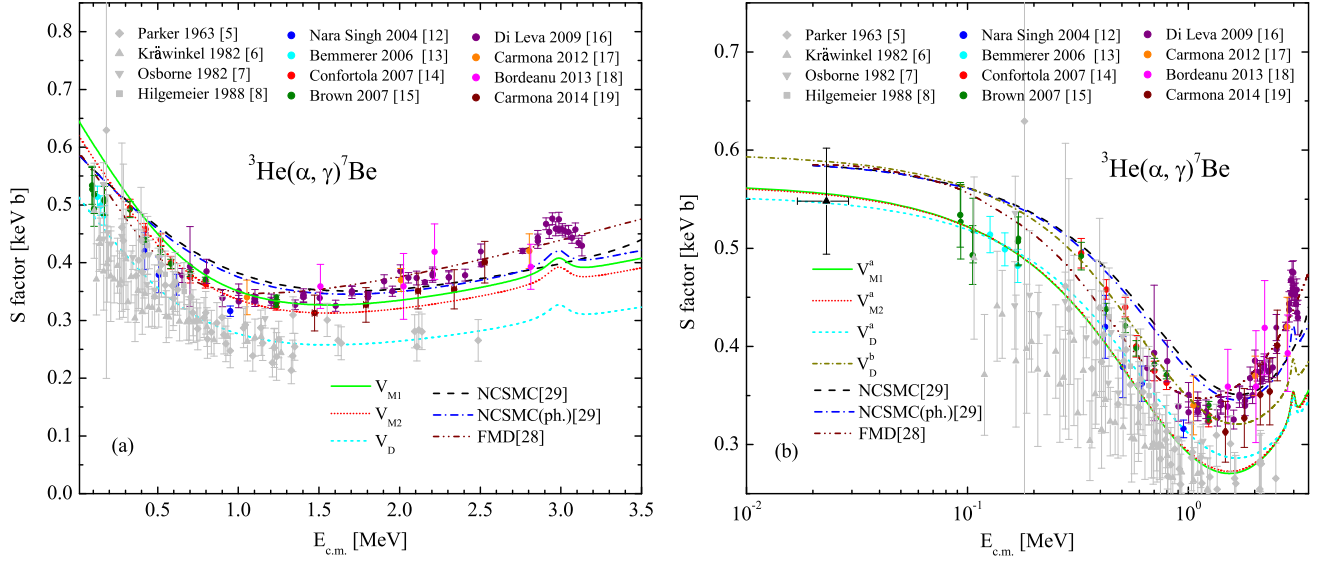}
\caption{Astrophysical S-factor for the $^{3}{\rm He}(\alpha,
\gamma)^{7}{\rm Be}$ synthesis reaction, estimated using different
potential models in comparison with available experimental data and
\emph{ab-initio} calculations. The right panel highlights the low-energy region.} \label{FIG6}
\end{figure}

The convergence of the astrophysical S-factor with respect to the
integration limit is demonstrated in Fig. \ref{FIG5} for the capture
process $^{3}{\rm He}(\alpha, \gamma)^{7}{\rm Be}$. As can be seen
from the figure, at low astrophysical energies the convergent
results are obtained with $R_{max}=$40 fm, while at higher energies
the convergence is reached already at $R_{max}=$20 fm.

In Fig. \ref{FIG6} we show the total astrophysical S-factor
for the  $^{3}{\rm He}(\alpha, \gamma)^{7}{\rm Be}$ capture
reaction. The left panel of the figure displays the astrophysical S-factor
of the process obtained by using the
$V_{M1}$, $V_{M2}$ and $V_D$ potentials in comparison with available
experimental data. As can be seen from the figure, the experimental
data is well reproduced at higher energies by the $V_{M1}$, $V_{M2}$
models, consistent with the NCSMC results \cite{doh16}. However,
these potential models  overestimate the data of the LUNA
collaboration \cite{bem06,confort07} at energies around 100 keV and
the newest data \cite{takacs15} at the Gamov peak
$S_{34}$(23$^{+6}_{-5}$ keV)=0.548$\pm$0.054 keV b. The reason is
that the energy dependence of the calculated astrophysical S-factor
is different from that of the microscopic NCSMC \cite{doh16}. The
potential model $V_D$ substantially underestimates the astrophysical
S-factor, although resulting energy dependence is similar to that obtained using
the $V_{M1}$ and $V_{M2}$ potentials.

As we already know, the E1-transitions $s_{1/2} \to p_{3/2}$ and
$s_{1/2} \to p_{1/2}$ play a dominant role in the capture process.
Therefore, the potentials $V_D$, $V_{M1}$, $V_{M2}$ can be modified
and their s-wave parameters $V_0$ and $\alpha$ can be adjusted to
the new data of the LUNA collaboration and the newest data at the
Gamov peak. As can be seen from right panel of Fig. \ref{FIG6}, the
potentials $V_D^a$, $V_{M1}^a$, $V_{M2}^a$ describe well the
astrophysical S-factor at low energies,
however they
have an increasing  tendency to underestimate the data above 0.5 MeV.
As discussed later, this underestimation is
not present for the mirror reaction.
A reason for the underestimation is that the potential model can not describe the
coupling to different inelastic channels, like $^6$Li$+p$ for the
process $^{3}{\rm He}(\alpha, \gamma)^{7}{\rm Be}$ or $^6$Li$+n$ for
the mirror capture process $^{3}{\rm H}(\alpha, \gamma)^{7}{\rm
Li}$. The main role in the coupling to inelastic channels is played
by the Coulomb forces and therefore the coupling should be important
for the first process. Also, tensor forces between valence nucleons
should play some role. These give rise to larger contribution of
higher partial waves of relative motion between valence nucleons.

The modified $V_D^b$ potential, whose s-wave parameters were fitted
to the upper limit of the newest data at the Gamov peak
\cite{takacs15}, yields a description of the experimental data in
both low and higher energy region with the same quality which is not
as good as for the  $V_D^a$, $V_{M1}^a$, $V_{M2}^a$ potentials.

As can be noted from the last figure, the energy dependence of the
astrophysical S-factor for the potential model slightly differs from
that resulting from the ab-initio study in the NCSMC \cite{doh16}
and is substantially different from energy dependence of the fermionic
molecular dynamics (FMD) model \cite{neff11} for the capture
process. The reason could be due to the fact that the experimental
s-wave phase shifts are not well reproduced by the NCSMC model.
A significantly different energy behavior of the FMD model and the potential approach could reflects the fact that the inputs  of the models are quite different.  The FMD model is a microscopic approach based on a realistic effective interaction that reproduces the nucleon-nucleon scattering data.  At the same time, the potential approach is based on the effective $\alpha + ^3$He ($^3$H) interaction potentials adjusted to the bound state properties of the $^7$Be nucleus and $\alpha + ^3$He scattering data.

The nodal positions of the s-wave scattering and the p-wave bound
state wave functions at small distances, which are due to
orthogonality to the Pauli forbidden states (two in s-wave and one
in each of the $p_{1/2}$ and $p_{3/2}$ partial waves), play a
crucial role in the description of the astrophysical S-factor. They
significantly affect the values of the overlap integral of the initial
and final state wave functions. A modification of the potential
parameters in the s- and p-waves is equivalent to shifting the nodal
positions of the s-wave scattering and p-wave bound state wave
functions. Thus the role of the Pauli forbidden states in the
capture process is similar to the important part they play in the
beta-decay process of the $^6${\rm He} halo nucleus
\cite{tur06,tur06a} and M1-transition of the $^6$Li(0$^+$)
\cite{tur07} isobar-analog state to the $\alpha-d$ two-body
continuum.

\subsection{Estimation of the astrophysical S-factor for the $^{3}{\rm H}(\alpha,
\gamma)^{7}{\rm Li}$ capture process}

As mentioned above the same $V_d$, $V_{M1}$, $V_{M2}$ potential
models and their modifications  $V_D^a$, $V_{M1}^a$, $V_{M2}^a$,
$V_D^b$ in the s-wave  are used for the study of the mirror capture
reaction $^{3}$H$(\alpha, \gamma)^{7}$Li. The Coulomb part of these
potentials, defined in Eq. (\ref{Coulomb}), is modified according to
the charge value of the $^3$H cluster $Z$=1. As demonstrated above, the
phase shifts in the $s_{1/2}$, $p_{1/2}$, $p_{3/2}$, $d_{3/2}$,
$d_{5/2}$, $f_{5/2}$ and $f_{7/2} $ partial waves, and the binding
energies $E_b(3/2^-)$=2.467 MeV and $E_b(3/2^-)$=1.990 MeV of the
bound states are well reproduced.

Partial contributions of the $E$1-, $E$2- and $M$1-transitions to
the astrophysical S-factor for the mirror $^{3}{\rm H}(\alpha,
\gamma)^{7}{\rm Li}$ reaction show the same behavior as for the
process $^{3}{\rm He}(\alpha, \gamma)^{7}{\rm Be}$.

\begin{figure}[htb]
\includegraphics[width=86mm]{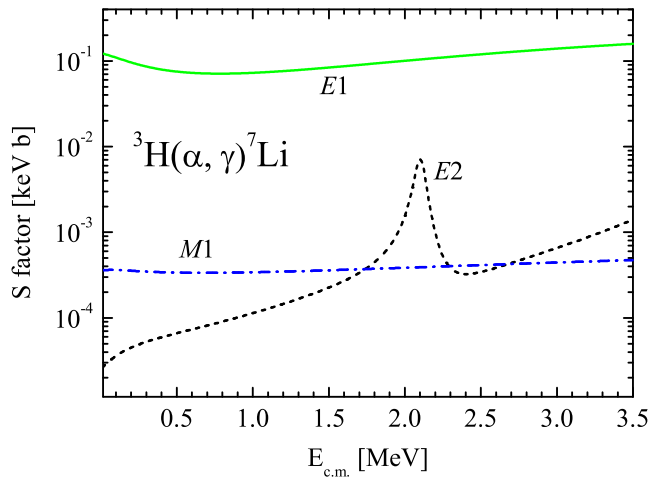}
\caption{Contributions of the $E$1, $E$2 and $M$1 transitions to the
astrophysical S-factor for the  $^{3}{\rm H}(\alpha, \gamma)^{7}{\rm
Li}$ synthesis reaction calculated with the  $V_{M1}$ potential.}
\label{FIG7}
\end{figure}

In Fig. \ref{FIG7} we show total contributions of the $E$1, $E$2 and
$M$1 transitions to the astrophysical S-factor for the $^{3}{\rm
H}(\alpha, \gamma)^{7}{\rm Li}$ synthesis reaction calculated with
the $V_{M1}$ potential model. As can be seen, the dominant role of
the E1-transition remains.

\begin{figure}[htb]
\includegraphics[width=178mm]{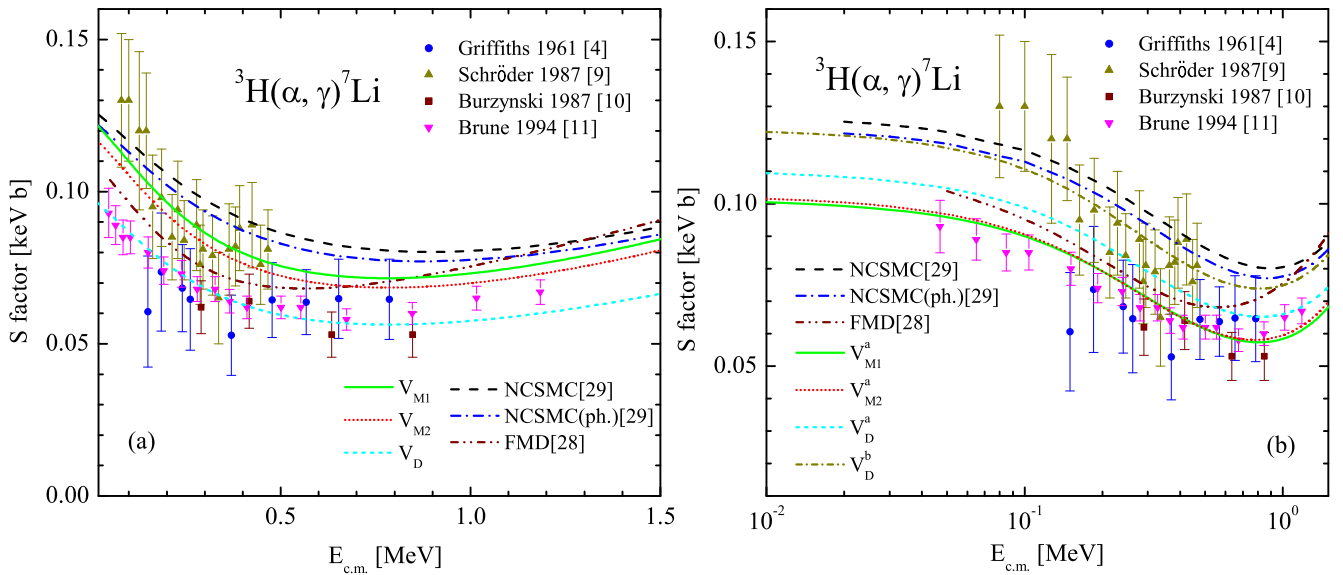}
\caption{Astrophysical S-factor for the $^{3}{\rm H}(\alpha,
\gamma)^{7}{\rm Li}$ synthesis reaction calculated with different
potentials in comparison with available experimental data and
\emph{ab-initio} calculations.} \label{FIG8}
\end{figure}

Finally, Figure \ref{FIG8} presents the total astrophysical S-factor
for the $^{3}{\rm H}(\alpha, \gamma)^{7}{\rm Li}$ reaction
calculated with the different potentials in comparison with
available experimental data. Since the latest data set
\cite{brune94} dates back to 1994, it is difficult to make any
conclusion on the experimental precision.
Nevertheless, one can see from the figure that the $V_D$,
$V_{M1}^a$, $V_{M2}^a$ potentials are more consistent with the
experimental data than the NCSMC \cite{doh16} and FMD \cite{neff11}
models. As in the case of the $^{3}{\rm He}(\alpha, \gamma)^{7}{\rm
Be}$ reaction, the energy dependence of the astrophysical S-factor
with the potential model is close to that with the NCSMC model but
substantially different from the FMD model. Even more importantly,
the $V_{M1}^a$, $V_{M2}^a$ potentials reproduce well both energy
dependence and normalization of the latest experimental data. This
means that a coupling to the inelastic channel $^6$Li$+n$ for the
mirror capture process $^{3}{\rm H}(\alpha, \gamma)^{7}{\rm Li}$ is
not important unlike the coupling to the $^6$Li$+p$ channel in the
$^{3}{\rm He}(\alpha, \gamma)^{7}{\rm Be}$ capture process. As was
noted earlier, the main role in the couplings to inelastic channels belongs
to Coulomb forces which are not present in the $^6$Li$+n$ channel.

\section{Conclusions}

The astrophysical $^{3}{\rm He}(\alpha,
\gamma)^{7}{\rm Be}$ and $^{3}{\rm H}(\alpha, \gamma)^{7}{\rm Li}$ direct capture processes   have been studied
in the two-body potential model. Central potentials of a simple
Gaussian form with the appropriate Coulomb part, which reproduce the
$\alpha-^3${\rm He} phase shifts in all the partial waves and binding
energies of the $^7$Be ground 3/2$^-$ and first excited 1/2$^-$
states, have been tested. It is important to note that the potentials, adjusted to the properties of the $^7$Be nucleus in
this way, were
able to reproduce the properties of the $^7$Li nucleus, phase shifts
in the partial waves and the binding energies of the ground 3/2$^-$
and first excited 1/2$^-$ states.

In addition, the potentials in the p-waves were adjusted to
reproduce the empirical values of the ANC for the $\alpha-^3$He,
extracted from the phase-shift analysis and alternatively, from the
analysis of the astrophysical S-factor available in the literature.

It has been shown that the E1-transition from the initial s-wave to
the final p-waves is strongly dominant in both capture reactions
considered in this work. On this basis we adjust the s-wave
potential to reproduce the  new data of the LUNA collaboration
around 100 keV and the latest data at the Gamov peak obtained on the
basis of the observed neutrino fluxes from the Sun,
$S_{34}$(23$^{+6}_{-5}$ keV)=0.548$\pm$0.054 keV b for the
astrophysical S-factor of the capture process $^{3}{\rm He}(\alpha,
\gamma)^{7}$Be. The resulting model describes well the experimental
data at low energies, however has an increasing tendency to underestimate the data
above 0.5 MeV. The underestimation could be due to the coupling
to the inelastic $^6$Li$+p$ channel, which can not be taken into account
in the developed potential model approach.
It is found that the energy dependence of the potential model is
slightly different from that of the microscopic no-core shell model
with continuum (NCSMC) and substantially differs from that of the
fermionic molecular dynamics (FMD) model.

It is also shown that the experimental data for the mirror
astrophysical $^{3}{\rm H}(\alpha, \gamma)^{7}{\rm Li}$ capture
reaction can be well described in the potential model. The
successful description of the data for the mirror process is
suggested to be due to a negligible role of the coupling to the
$^6$Li$+n$ inelastic channel in which the Coulomb forces are not
present.

In conclusion, the $V_{M1}^a$, $V_{M2}^a$ potential models which
were fitted to the new data of the LUNA collaboration for the
astrophysical S-factor of the $^{3}{\rm He}(\alpha, \gamma)^{7}{\rm
Be}$ capture process by the modifying the s-wave $\alpha-^3$He
nuclear interaction potential describe well this capture process in the
BBN energy region (180-400 keV). Additionally, they yield very good
description of the latest experimental data for the $^{3}{\rm
H}(\alpha, \gamma)^{7}{\rm Li}$ mirror capture process. From the
beginning, these models describe well bound state (binding energies and
ANC) and scattering state (phase shifts) properties of both
$\alpha+^3$He and $\alpha+^3$H systems.

\begin{acknowledgments}
A.S.K acknowledges support from the Australian Research Council and
partial support from the U.S.\ National Science Foundation
under Award No. PHY-1415656.
We would like to thank J. Dohet-Eraly for providing us with the results of Ref. \cite{doh16} for the astrophysical S-factor in a tabulated form.
\end{acknowledgments}

\end{document}